\documentclass[pra,aps,twocolumn,superscriptaddress]{revtex4}
\usepackage{hyperref}
\usepackage{bm}
\usepackage{epsfig}
\usepackage{amssymb}
\usepackage{graphicx}

\usepackage{amsmath}
\usepackage{epsfig}

\begin{document}
\title{Distribution of continuous-variable entanglement by separable Gaussian states}
\author{Ladislav Mi\v{s}ta, Jr.}
\affiliation{Department of Optics, Palack\' y University, 17.
listopadu 50,  772~07 Olomouc, Czech Republic} \affiliation{School
of Physics and Astronomy, University of St. Andrews, North Haugh,
St. Andrews, KY16 9SS, UK}
\author{Natalia Korolkova}
\affiliation{School of Physics and Astronomy, University of St.
Andrews, North Haugh, St. Andrews, KY16 9SS, UK}

\date{\today}

\begin{abstract}
Entangling two systems at distant locations using a {\it separable} mediating ancilla
is a counterintuitive phenomenon
proposed for qubits by T. Cubitt {\it et al}. [Phys. Rev. Lett.
{\bf 91}, 037902 (2003)].
We show that such entanglement distribution is possible with Gaussian states,
using a certain three-mode fully separable
mixed Gaussian state and linear optics elements readily available
in experiments. Two modes of the state become entangled by
sequentially mixing them on two beam splitters,
while the third one remains separable in all stages of the protocol.
\end{abstract}

\maketitle

Quantum entanglement is a striking property of composite quantum
systems that lies at the heart of the fundamental quantum
information protocols such as quantum teleportation
\cite{Bennett_93} or quantum cryptography \cite{Ekert_91}.
The typical scenario involves two parties in distant laboratories, Alice holding
the quantum system $a$ and Bob in possession of the system $b$.
They need to establish a quantum channel between their remote location in the form of
the shared entangled state, which cannot be prepared merely by local operations on
systems $a$ and $b$ and classical communication between Alice and Bob \cite{Werner_89}.
Having only separable systems at hand and not having a possibility
to meet each other in one place, Alice and Bob can entangle
their distant quantum systems only by employing
another ancillary quantum system $c$, which first couples with the system $a$, then is send over to the remote location where it interacts with $b$. This is  a required global operation that facilitates entanglement between $a$ and $b$. Remarkably, $a$ can be entangled with $b$ by sending the ancilla
$c$ that becomes never entangled with the subsystem $(ab)$
\cite{Cubitt_03}. In \cite{Cubitt_03} the counterintuitive
effect of entanglement distribution by separable ancilla
was studied in the context of finite-dimensional systems.

In this paper we show how to turn this idea into a practical concept.
We consider infinite-dimensional quantum
systems, e.g., light modes.
We propose a feasible three-step protocol
where two distant separable modes $A$ and $B$ become entangled
after interacting stepwise with the third mode $C$.
At any stage of the protocol,
the mode $C$ is separable from the subsystem $(AB)$.
Our scheme relies entirely on Gaussian states and the
challenging nonlinear controlled-NOT gates of
the previous idea \cite{Cubitt_03}
are replaced by simple beam splitters.
Therefore,
the protocol can be implemented with Gaussian states
and operations that are currently available in the laboratory.
Moreover, the proposed protocol allows a more simple
deterministic distribution of entanglement than the previous
qubit protocol that requires an additional operation on systems
$b$ and $c$ on Bob's side \cite{Cubitt_03}.

\begin{figure}[!]
\centerline{\psfig{width=6.5cm,angle=0,file=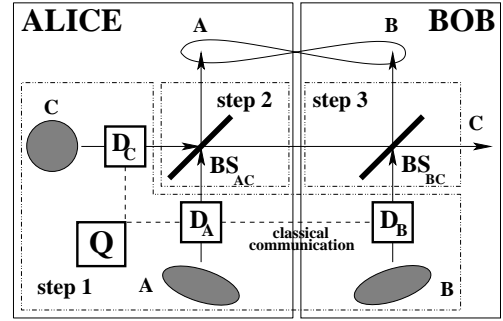}}
\caption{Scheme of the protocol for distribution of
continuous-variable entanglement by separable Gaussian states.
Step 1: LOCC preparation of a fully separable Gaussian state of
three modes $A$, $B$ and $C$. Step 2 entangles mode $A$ with a pair
of modes $(BC)$. Step 3 entangles mode $A$ with mode $B$. Mode $C$
remains separable from the pair of modes $(AB)$ in all steps. $BS_{{AC},{BC}}$
denote balanced beam splitters, $D_{A},D_{B}$, $D_{C}$ are local
displacements distributed according to the Gaussian distribution with correlation matrix $Q$.} \label{fig1}
\end{figure}
Our protocol is schematically depicted in Fig.~\ref{fig1}. The aim
of the protocol is to entangle mode $A$ in Alice's laboratory with
separable mode $B$ in Bob's distant laboratory by sending a
separable mediating ancillary mode $C$ from Alice to Bob. For pure quantum states
this is not possible \cite{Cubitt_03}. Therefore, Alice and Bob have to construct
by local operations and classical communication (LOCC) a suitable {\it mixed}
fully separable Gaussian state of three modes $A$, $B$ and $C$.
The engineering of such a state is the aim of step 1 and
represents the most challenging part of the problem. Alice and Bob
start with three pure single-mode Gaussian states. They prepare modes
$A$ and $B$ in the same momentum-squeezed vacuum states and rotate
them clockwise and anticlockwise, respectively, by the same suitable
angle. The ancillary mode $C$ is initially on Alice's
side and is in a vacuum state. Alice and Bob then displace locally
the three modes by random correlated displacements $D_{A},D_{B}$
and $D_{C}$ with Gaussian distribution characterized
by a correlation matrix $Q(x)$ specified below.
This procedure generates the desired three-mode
mixed Gaussian state with separability properties tunable by changing the
parameter $x$.

The actual entanglement distribution commence in step 2. By mixing modes
$A$ and $C$ on a balanced beam splitter $BS_{AC}$ Alice entangles mode $A$
with the pair of modes $(BC)$ while mode $B$ is separable from $(AC)$ and
mode $C$ is separable from $(AB)$. Next, she sends mode $C$ to Bob.
In step 3 Bob mixes modes $B$ and $C$ on a balanced beam splitter $BS_{BC}$
finally entangling $A$ and $B$, while $C$ still remains separable
from $(AB)$.

The modes $A$, $B$, and $C$  are described by three pairs of
canonically conjugate quadrature operators $x_{j},p_{j}$,
$j=A,B,C$. The operators satisfy the canonical commutation rules
that can be compactly expressed as
$[\xi_{j},\xi_{k}]=-i\Omega_{jk}$, where
$\xi=(x_{A},p_{A},x_{B},p_{B},x_{C},p_{C})^{T}$ is the vector of
quadratures and $\Omega=\oplus_{i=1}^3 J$ is the symplectic
matrix, where $J=-i\sigma_y$ ($\sigma_y$ denotes the $y$ Pauli
matrix). Quantum states of three-mode system can be represented in
phase space by the Wigner function \cite{Wigner_32} of six real
variables and Gaussian states are defined as those having a
Gaussian-shaped Wigner function. Any three-mode Gaussian state $\rho$ is
therefore fully characterized by the vector of first moments
$\bar{\xi}=\mbox{Tr}(\rho\xi)$, that we assume to be zero, and by
the $6\times 6$ real symmetric covariance matrix (CM) $\gamma$
with elements $\gamma_{jk}=\mbox{Tr}\left(\rho\{\xi_{j}-\bar
{\xi}_{j}\openone,\xi_{k}-\bar{\xi}_{k}\openone\}\right)$,
$j,k=1,\ldots,6$, where $\{A,B\}\equiv AB+BA$.

\paragraph*{Preparation of the three-mode fully separable state.}
We start with the three-mode Gaussian
state, which CM is composed from the CM of an entangled state  $\gamma_{AB}$
\begin{eqnarray}\label{gammaAB}
\gamma_{AB}=\left(\begin{array}{cccc}
e^{2d}a & 0 & -e^{2d}c & 0 \\
0 & e^{-2d}a & 0 & e^{-2d}c \\
-e^{2d}c & 0 & e^{2d}a & 0 \\
0 & e^{-2d}c & 0 & e^{-2d}a \\
\end{array}\right)
\end{eqnarray}
and a noise term in the form of a nonnegative multiple of a positive semidefinite matrix $P\equiv q_1q_1^{T}+q_2q_2^{T}$:
\begin{eqnarray}\label{gamma1}
\gamma_{1}(x)=\gamma_{AB}\oplus\openone_{C}+x(q_1q_1^{T}+q_2q_2^{T}),
\end{eqnarray}
where $x\geq0$. The parameters involved in the CM ({\ref{gammaAB}}) are given by
$a=\cosh(2r)$, $c=\sinh(2r)$ and we assume $d\geq r>0$. This is a two-mode squeezed vacuum state with the
squeezing parameter $r$ with modes $A$ and $B$ squeezed, in addition, by local squeezing operations
$S_{A}=S_{B}=\mbox{diag}(e^{d},e^{-d})$. Our design of the noise term
in Eq.~(\ref{gamma1}) is inspired by the method \cite{Giedke_01} used to
construct various three-mode entangled Gaussian states and
$q_1, q_2$ read
\begin{eqnarray}\label{q1q2}
q_1&=&(0,\sin\phi,0,-\sin\phi,\sqrt{2},\sqrt{2})^{T},\nonumber\\
q_2&=&(\cos\phi,0,\cos\phi,0,\sqrt{2},\sqrt{2})^{T},\\
\tan\phi &=& e^{-2r}\sinh(2d)+\sqrt{1+e^{-4r}\sinh^{2}(2d)} \nonumber
\end{eqnarray}
with $\sin\phi,\cos\phi>0$. This additional noise is chosen such that for
sufficiently large $x$ the CM (\ref{gamma1}) describes a fully separable state.

The state described by CM (\ref{gammaAB}) can be naturally prepared
by mixing on a balanced beam splitter 
$U_{AB}$ \cite{BS} modes $A$ and $B$, each in a pure momentum-squeezed vacuum state with
the variances of the position quadratures
$\langle(\Delta x_{A})^{2}\rangle=e^{2(d-r)}$ and
$\langle(\Delta x_{B})^{2}\rangle=e^{2(d+r)}$ respectively. The entire three-mode state with CM
(\ref{gamma1}) then can be created by adding a vacuum mode $C$ with CM
$\openone_{C}$ to the CM $\gamma_{AB}$ and performing local random
correlated displacements of modes $A$, $B$ and $C$ distributed
with Gaussian distribution with correlation matrix $xP$ \cite{Werner_01}.
Making use of the criterion
of full separability for three-mode Gaussian states
\cite{Giedke_01} one then finds that for all $x\geq x_{\rm sep}$, where
\begin{equation}\label{xsep}
x_{\rm sep}=\frac{2\sinh(2r)}{\delta},
\end{equation}
where $\delta=e^{2d}\sin^{2}\phi+e^{-2d}\cos^{2}\phi$, the CM
(\ref{gamma1}) describes a fully separable state. However, although
the state is fully separable the way of its preparation described
above is not suitable for our purposes. Namely, it is not
prepared by LOCC but instead requires Alice and Bob to meet to implement the beam
splitting operation $U_{AB}$ on their modes $A$ and $B$.

Still the CM (\ref{gamma1}) corresponds to a fully separable state
and therefore there exists a recipe how to create this state by LOCC. The
recipe is based on the three-mode separability criterion \cite{Giedke_01,Werner_01}
according to which a three-mode Gaussian state with CM $\gamma_{1}(x)$ is
fully separable iff there exist single-mode CMs
$\gamma_{A}$, $\gamma_{B}$ and $\gamma_{C}$ such that
\begin{equation}\label{fullseparability}
Q(x)\equiv\gamma_{1}(x)-\gamma_{A}\oplus\gamma_{B}\oplus\gamma_{C}\geq0.
\end{equation}
Interestingly, such single-mode CMs can be indeed found for $x\geq x_{\rm sep}$ in the form
\begin{eqnarray}\label{single-mode}
\gamma_{A,B} & = & \left(\begin{array}{cc}
\alpha+\beta & \mp\tau \\
\mp\tau & \alpha-\beta \\
\end{array}\right),\quad \gamma_{C}=\openone,
\end{eqnarray}
where
\begin{eqnarray}\label{xyz}
\alpha&=&\frac{e^{-2r}}{2\delta}\left[e^{4r}+\cosh(4d)-\sinh(4d)\cos(2\phi)\right],\nonumber\\
\beta&=&\frac{e^{-2r}}{2\delta}\left\{\left[e^{4r}-\cosh(4d)\right]\cos(2\phi)+\sinh(4d)\right\},\nonumber\\
\tau&=&\frac{\sinh(2r)}{\delta}\sin(2\phi),
\end{eqnarray}
and the parameters satisfy the purity condition $\alpha^{2}=\beta^{2}+\tau^{2}+1$.
The CM $\gamma_{C}$ represents a vacuum state. The CM $\gamma_{A}$ ($\gamma_{B}$) corresponds to the pure momentum-squeezed vacuum state with squeezing parameter $s=\frac{1}{2}\ln\left(\alpha+\sqrt{\alpha^{2}-1}\right)$
rotated clockwise (anticlockwise) by the phase
$\theta=\arctan\left(\sqrt{\frac{\sqrt{\alpha^{2}-1}-\beta}{\sqrt{\alpha^{2}-1}+\beta}}\right)$.

It remains to show that the matrix $Q(x)$ is positive semidefinite for $x\geq x_{\rm sep}$.
It is sufficient to show that for $x=x_{\rm sep}$ since if $Q(x_{\rm sep})\geq 0$, then
$Q(x)=Q(x_{\rm sep})+(x-x_{\rm sep})P\geq 0$ for all $x\geq x_{\rm sep}$
because $(x-x_{\rm sep})P$ is also positive semidefinite. To obtain the eigenvalues of the matrix $Q(x_{\rm sep})$, we will calculate the eigenvalues of the matrix $U_{AB}Q(x_{\rm sep})U_{AB}^{T}$, which possesses the same eigenvalues. They read explicitly as
$\lambda_{1,2,3,4}=0$, $\lambda_{5}=9x_{\rm sep}$ and $\lambda_{6}=\left(e^{4d}\sin^{2}\phi+e^{-4d}\cos^{2}\phi\right)x_{\rm sep}$.
All of the eigenvalues are nonnegative and therefore the matrix $Q(x)$ for
$x\geq x_{\rm sep}$ is indeed positive semidefinite.

Creation of the fully
separable state with CM $\gamma_{1}(x)$, where $x\geq x_{\rm sep}$, is now straightforward
\cite{Werner_01}. Initially, Alice prepares in her laboratory mode $A$ in a pure
single-mode squeezed state with CM $\gamma_{A}$ and the ancillary mode $C$ in the
vacuum state. Similarly, Bob prepares the mode $B$ in a pure single-mode
squeezed state with CM $\gamma_{B}$. In the next step, Alice and Bob displace locally
their modes by random correlated displacements distributed according to the Gaussian
distribution with correlation matrix $Q(x)$. As a result, they prepare by LOCC a
three-mode fully separable Gaussian state with CM (\ref{gamma1}). For the sake of
simplicity here and in what follows we do not write explicitly the dependence
of CMs on the parameter $x$ and we implicitly assume that $x\geq x_{\rm sep}$.

\paragraph*{Entanglement distribution.}
In step 2 Alice superimposes modes $A$ and $C$ of a fully separable
state described by the CM $\gamma_1$ on a balanced beam splitter $U_{AC}$ \cite{BS}
that transforms the CM as
\begin{equation}\label{gamma2}
\gamma_{2}=U_{AC}\gamma_{1}U_{AC}^{T}.
\end{equation}
Apparently, the CM is separable with respect to partition
$B-(AC)$. More interestingly, mode $C$ can remain separable from
the subsystem $(AB)$ if we choose the parameters
 $d,r$ and $x$ properly. To prove this, we apply to CM (\ref{gamma2}) the
separability criterion based on the symplectic invariants
\cite{Serafini_06}. The criterion utilizes the matrix $\gamma_{2}^{T_{C}}=\Lambda_{C}\gamma_{2}\Lambda_{C}$,
where $\Lambda_{C}=\mbox{diag}(1,1,1,1,1,-1)$, that describes CM
$\gamma_{2}$ after partial transposition
with respect to the mode $C$ \cite{Peres_96,Horodecki_96,Simon_00,Duan_00,Werner_01,Giedke_01}.
The matrix $\gamma_{2}^{T_{C}}$ has three symplectic invariants denoted $I_1,I_2$ and
$I_3=\mbox{det}(\gamma_{2})$ that can be obtained as coefficients
of the characteristic polynomial of the matrix
$\Omega\gamma_{2}^{T_{C}}$, i.e.
$\mbox{det}(\Omega\gamma_{2}^{T_{C}}-y\openone)=y^{6}+I_1y^{4}+I_2y^{2}+I_3.$
According to the criterion mode $C$ is separable from modes $(AB)$ iff
\begin{equation}\label{Sigma}
\Sigma\equiv I_1-I_2+I_3-1\geq0
\end{equation}
holds \cite{strictly}. Calculating the invariants using the equations above we
arrive at the following simple expression:
\begin{equation}\label{quadratic}
\Sigma=x(ux+v),
\end{equation}
where $u$ and $v$ are complex functions of parameters $d$ and $r$
given elsewhere \cite{Mista_prep}. In this paper we are interested in
demonstrating the possibility of entangling $A$ and $B$, while
keeping $C$ separable. So we show that for particular values of
parameters $d$ and $r$ we get $\Sigma>0$ and hence CM
$\gamma_{2}$ is separable for $x$ larger than a certain threshold
value $x_{\rm th}$. Eq.~(\ref{quadratic}) determines a parabola in
the $(x,\Sigma)$ plane that intersects the $x$-axis in the origin.
Taking $e^{2(d-r)}=3/2$ and $e^{2(d+r)}=2$ (which corresponds to
$e^{-2s}\approx0.6387$ and $\theta\approx5.73^{\circ}$) we get $u>0$ and $v<0$
and the parabola is oriented upwards. Making use of
Eqs.~(\ref{xsep}) and (\ref{quadratic}), the threshold value
$x_{\rm th}=-v/u\approx1.04>x_{\rm sep}\approx0.2043$, and hence
for $x>x_{\rm th}$ we have $\Sigma>0$ and the CM (\ref{gamma2}) is
separable with respect to the partition $C-(AB)$.

The protocol is finalized in the step 3. After receiving mode $C$ from Alice,
Bob superimposes this mode with his mode $B$ on another balanced
beam splitter $BS_{BC}$ \cite{BS}. The CM of the resulting state reads
\begin{equation}\label{gamma3}
\gamma_{3}=U_{BC}\gamma_{2}U_{BC}^{T}.
\end{equation}
Remarkably, $\gamma_3$ exhibits entanglement between modes $A$ and $B$
whereas mode $C$ remains separable from $(AB)$.

To verify entanglement between modes $A$ and $B$ we express the
two-mode CM $\gamma_{3,AB}$ of the reduced state of modes $A$ and $B$
in the block form
\begin{eqnarray}\label{gamma-block}
\gamma_{3,AB} & = & \left(\begin{array}{cc}
{\cal A} & {\cal C} \\
{\cal C}^{T} & {\cal B} \\
\end{array}\right)
\end{eqnarray}
with the submatrices ${\cal A}$, ${\cal B}$, and ${\cal C}$ of the form:
\begin{eqnarray}
\label{calABC}
{\cal C}& = & \left(\begin{array}{cc}
c_{+}+(g_1h_1-\frac{1}{\sqrt{2}})x & -(h_{0}+\frac{g_{1}}{\sqrt{2}})x \\
(h_{1}-\frac{g_{0}}{\sqrt{2}})x  & c_{-}-(g_{0}h_{0}+\frac{1}{\sqrt{2}})x \\
\end{array}\right),\nonumber
\end{eqnarray}
\begin{eqnarray}
{\cal A}& = & \left(\begin{array}{cc}
a_{+}+(g_1^{2}+1)x & (g_{0}+g_{1})x \\
(g_{0}+g_{1})x & a_{-}+(g_0^{2}+1)x \\
\end{array}\right),\nonumber\\
{\cal B}& = & \left(\begin{array}{cc}
b_{+}+(h_1^{2}+\frac{1}{2})x & \frac{h_{0}-h_{1}}{\sqrt{2}}x \\
\frac{h_{0}-h_{1}}{\sqrt{2}}x  & b_{-}+(h_0^{2}+\frac{1}{2})x \\
\end{array}\right),
\end{eqnarray}
where $a_{\pm}=(e^{\pm 2d}a+1)/2$; $b_{\pm}=[e^{\pm
2d}(3a\mp2\sqrt{2}c)+1]/4$; $c_{\pm}=[e^{\pm
2d}(a\mp\sqrt{2}c)-1]/2\sqrt{2}$;
$g_{j}=1+\sin(\phi+j\frac{\pi}{2})/\sqrt{2}$;
$h_{j}=[\sqrt{2}-(-1)^{j}]\sin(\phi+j\frac{\pi}{2})/2+(-1)^{j}/\sqrt{2}$,
$j=0,1$. The entanglement of CM $\gamma_{3,AB}$ can be proved if we calculate
the so called symplectic eigenvalues \cite{Williamson_36} of the matrix $\gamma_{3,AB}^{T_{B}}=\Lambda_{2,B}\gamma_{3,AB}\Lambda_{2,B}$,
where $\Lambda_{2,B}=\mbox{diag}(1,1,1,-1)$ \cite{Vidal_02}. The matrix has two symplectic
eigenvalues $\nu,\nu'$ that can be computed from the eigenvalues of
the matrix $\Omega_{2}\gamma_{3,AB}^{T_{B}}$, where $\Omega_{2}=\oplus_{i=1}^2 J$ and are equal to
 $\{\pm i\nu,\pm i\nu'\}$ \cite{Vidal_02}. The mode $A$ is entangled with the mode
$B$ iff $\nu<1$ or $\nu'<1$. The lower symplectic eigenvalue of the matrix
$\gamma_{3,AB}^{T_{B}}$ can be expressed as \cite{Vidal_02}
\begin{eqnarray}\label{ABsymplectic}
\nu=\sqrt{\frac{\kappa-\sqrt{\kappa^2-4\mbox{det}(\gamma_{3,AB})}}{2}},
\end{eqnarray}
where $\kappa=\mbox{det}({\cal A})+\mbox{det}({\cal
B})-2\mbox{det}({\cal C})$ and  ${\cal A}$, ${\cal B}$, ${\cal C}$
are defined in Eq.~(\ref{calABC}). By the same token as in step 2, we
revert to the case $e^{2(d-r)}=3/2$, $e^{2(d+r)}=2$. Taking
$x=1.041>x_{\rm th}$, we obtain an exact expression for $\nu$ in
terms of square roots that approximately equals to
$\nu\approx0.9571<1$, which is a clear evidence of the
entanglement between modes $A$ and $B$. Note, that by measuring a
quadrature $(x_{C}+p_{C})/\sqrt{2}$ on mode $C$ the eigenvalue
$\nu$ can be further reduced (i.e. entanglement can be increased \cite{Adesso_04}) to
$\nu_{\rm m}=0.9421$.

Finally, we have to show that the ancillary mode $C$ is separable
from the two-mode subsystem $(AB)$. We use again the simplectic invariants criterion of
\cite{Serafini_06}. Analogous to step 2, we calculate the characteristic polynomial
of the matrix $\Omega\gamma_3^{T_C}$ and find three symplectic
invariants $J_{1}$, $J_{2}$ and $J_{3}=\mbox{det}(\gamma_{3})$, which have to obey the condition
$\tilde \Sigma=J_1-J_2+J_3-1=x(wx+z)\geq0$ (c.f.
(\ref{Sigma}), (\ref{quadratic})). The parameters $w$ and $z$ are again complex
functions of the parameters $d$, $r$ 
\cite{Mista_prep}. Assuming $e^{2(d-r)}=3/2$, $e^{2(d+r)}=2$ and
$x=1.041$, we obtain $\tilde \Sigma\approx0.3957>0$. Thus the mode
$C$ is separable from $(AB)$.

In an experiment, verification of entanglement of modes $A$ and $B$ in step 3
can be done by measuring the entire CM $\gamma_{3,AB}$ similarly as in \cite{DiGuglielmo_07} and
applying Simon's \cite{Simon_00} or Duan's \cite{Duan_00} separability criterion.
The separability of mode $C$ from the pair of modes $(AB)$ in steps 2 and 3 can
be proved by measuring the three-mode CMs $\gamma_{2}$ and $\gamma_{3}$. For
Gaussian states one then can use the positive partial transposition criterion \cite{Peres_96,Horodecki_96}
that is sufficient for separability of these $1\times 2$-mode systems \cite{Werner_01}.

\begin{figure}
\centerline{\psfig{width=8.0cm,angle=0,file=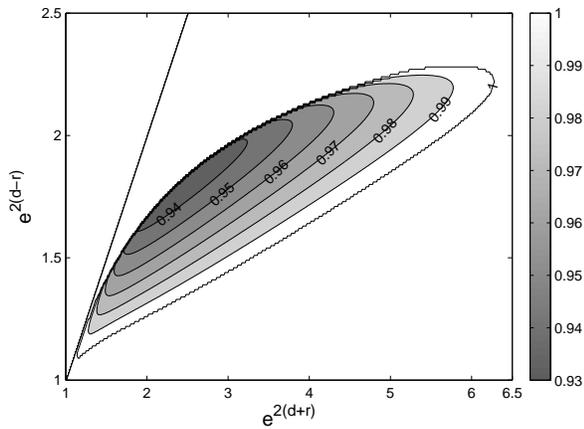}} \caption{
Performance of our protocol in dependence of the variances
$\langle(\Delta x_{A})^{2}\rangle=e^{2(d-r)}$ and
$\langle(\Delta x_{B})^{2}\rangle=e^{2(d+r)}$. Gray-scale region: all assumptions of the protocol are satisfied and $\nu<1$.
White region to the right of the slash: either $\nu\geq1$ or $C$ is entangled with
$(AB)$ in some stage of the protocol. The contour lines display the values of
symplectic eigenvalue $\nu$.}\label{fig2}
\end{figure}

As a numerical evidence of the robustness of the protocol, we obtained eigenvalue
$\nu=0.9787$ for $\gamma_1+2\times10^{-2}\openone$
corresponding to the initial CM perturbed by a weak isotropic noise. Numerical analysis also
verifies the performance of the protocol for a broad range of
variances $\langle(\Delta x_{A,B})^{2}\rangle=e^{2(d\mp r)}$ depicted
by a gray region in Fig.~\ref{fig2}. 

In conclusion, we have demonstrated the possibility to distribute
entanglement without sending entanglement in infinite-dimensional
systems. Remarkably, one can entangle two distant modes by
a separable mode using experimentally feasible Gaussian states and operations
involving single-mode squeezed states, correlated displacements and
beam splitters, dispensing with the CNOT gates of the qubit case. The
distributed entanglement is distillable \cite{Giedke_01b} and
therefore can be used for quantum communication. In contrast with
two qubits, two light modes can be entangled deterministically
even without any additional operation on Bob's
system and ancilla beyond step 3 (cf. \cite{Cubitt_03}).
Furthermore, we have elaborated the procedure to design three-mode
Gaussian states with desired separability and noise properties.
Together with the possibility to distribute distillable
continuous-variable entanglement without sending it through
the channel, it prepares the ground for better understanding
and engineering of optical quantum networks, continuous-variable
cryptography and other entanglement-based communication
protocols using light modes and/or atomic ensembles.
The support of the EU project
COVAQIAL (FP6-511004) under STREP and the Czech Ministry of
Education (Grant Nos. MSM 6198959213 and LC06007) is acknowledged.


\begin{thebibliography}{99}
\bibitem{strictly} Strictly speaking, $\Sigma>0$ suffices for
separability of CM $\gamma_2$ as there are entangled CMs giving
$\Sigma=0$ \cite{Serafini_06}.
\end{thebibliography}
\end{document}